\documentclass[
    amsmath,amssymb,
    superscriptaddress,
    reprint, prl
   ]{revtex4-1}
\newcommand{\expect}[1]{
    \ensuremath{\langle{#1}\rangle}
}

\newcommand{\Tr}{\text{Tr}}
\newcommand{\poly}{\mathrm{poly}}

\usepackage{graphicx}
\usepackage{color}
\usepackage{dcolumn}
\usepackage{siunitx}
\usepackage{bm}
\usepackage{braket}

\begin{document}


\title[]{Methodology for replacing indirect measurements with direct measurements}

\author{Kosuke Mitarai}
\email{u801032f@ecs.osaka-u.ac.jp}
\affiliation{Graduate School of Engineering Science, Osaka University, 1-3 Machikaneyama, Toyonaka, Osaka 560-8531, Japan.}
\affiliation{QunaSys Inc, High-tech Hongo Building 1F, 5-25-18 Hongo, Bunkyo, Tokyo 113-0033, Japan}
\author{Keisuke Fujii}
\email{fujii.keisuke.2s@kyoto-u.ac.jp}
\affiliation{Graduate School of Science, Kyoto University, Kitashirakawa Oiwake-cho, Sakyo-ku, Kyoto 606-8302, Japan.}
\affiliation{JST, PRESTO, 4-1-8 Honcho, Kawaguchi, Saitama 332-0012, Japan}
   
\date{\today}

\begin{abstract}
    In quantum computing, the indirect measurement of unitary operators such as the Hadamard test plays a significant role in many algorithms.
    However, in certain cases, the indirect measurement can be reduced to the direct measurement, where a quantum state is destructively measured.
    Here we investigate in what cases such a replacement is possible and develop a general methodology for trading an indirect measurement with sequential direct measurements.
    The results can be applied to construct quantum circuits to evaluate the analytical gradient, metric tensor, Hessian, and even higher order derivatives of a parametrized quantum state.
    Also, we propose a new method to measure the out-of-time-order correlator based on the presented protocol.
    Our protocols can reduce the depth of the quantum circuit significantly by making the controlled operation unnecessary and hence are suitable for quantum-classical hybrid algorithms on near-term quantum computers.
\end{abstract}

\pacs{Valid PACS appear here}
\maketitle

\begin{figure}
    \includegraphics[width=0.6\linewidth]{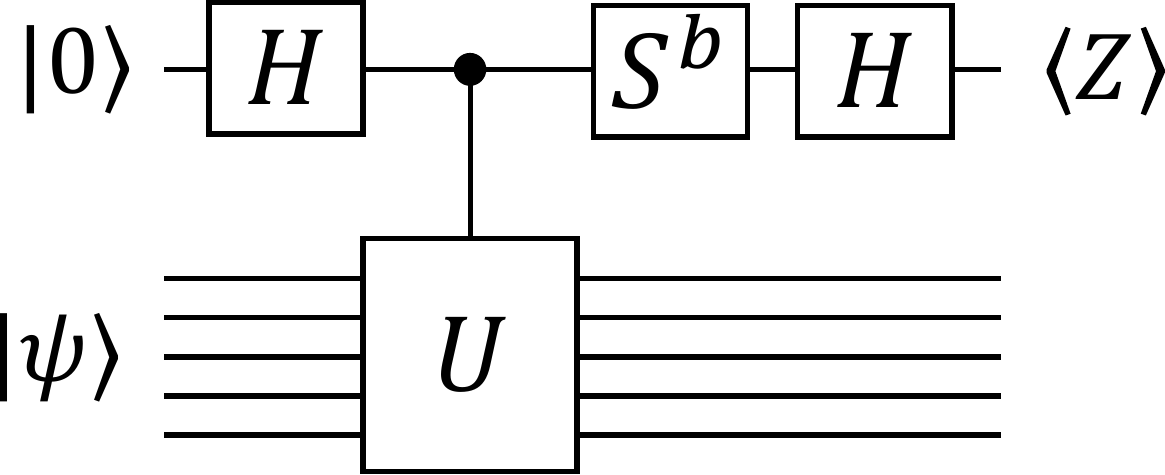}
    \caption{\label{fig:Hadamard_test} Simplest Hadamard test. In the figure, $b\in \{0,1\}$ and $U$, $H$, $S$ are an arbitrary quantum gate, the Hadamard gate, and $e^{-i\pi Z/4}$, respectively. When $b = 0$, $\expect{Z} = \mathrm{Re}\braket{\psi_{\text{in}}|U|\psi_{\text{in}}}$ and when $b=1$, $\expect{Z} = \mathrm{Im}\braket{\psi_{\text{in}}|U|\psi_{\text{in}}}$.}
\end{figure}

\noindent\textit{Introduction -}
The output from quantum computation is measured in two ways; indirect and direct measurement of observables.
In the former, the measured quantum state is not completely destructed, whereas, in the latter, the state collapses to the basis on which we perform the measurement.
The simplest and important protocol for the indirect method is the Hadamard test (Fig.~\ref{fig:Hadamard_test}).
In the Hadamard test, we add an ancillary qubit and apply a controlled unitary gate, a unitary $U$ to a target quantum state $\ket{\psi}$ conditioned on the ancilla being $\ket{0}$ or $\ket{1}$ to measure expectation value of $\braket{\psi|U|\psi}$ as the $Z$ expectation value of the ancilla.
This measurement allows us to reuse the state $(I\pm U)\ket{\psi}/\sqrt{2}$ after the measurement, which is the property exploited in algorithms like iterative phase estimation \cite{Knill2007, Dobsicek2007}.

Such indirect approaches can achieve the precision of $\epsilon$ in $O(1/\epsilon)$ time.
However, the implementation of the controlled-$U$ gate can be a hard task especially for so-called noisy intermediate scale quantum (NISQ) \cite{Preskill2018} devices.
In fact, direct measurements can be satiable when only an expectation value of an observable is required.
A famous example is estimation of energy expectation values in the variational quantum eigensolver (VQE) \cite{Peruzzo2013}, which is one of the most promising applications of NISQ devices.
The time required to achieve the precision of $\epsilon$ is $O(1/\epsilon^2)$ in this approach, which is in contrast with the indirect approach \cite{Wecker2015, Wang2018}.

Another example, which replaces the indirect approach with the direct one, is the destructive swap test \cite{Garcia-Escartin2013}.
It is a direct version of the swap test \cite{Buhrman2001} which measures the overlap $|\braket{\psi|\varphi}|^2$ between two quantum states $\ket{\psi}$ and $\ket{\varphi}$.
After a while from the first proposal \cite{Garcia-Escartin2013}, it has been rediscovered by machine learning approach \cite{Cincio2018a}, and now it is utilized for applications of NISQ devices \cite{Higgott2018,Endo2018,LaRose2018}. 
Ref.~\cite{Subasi2018} has proposed to use the destructive swap test to measure $|\braket{\psi|U|\psi}|^2$ for arbitrary \(U\) by substituting $\ket{\varphi}$ with $U\ket{\psi}$, and also extended the protocol to measure the quantity $|\braket{\psi|P|\varphi}|^2$, where $P$ is a qubit-permutation operator, which can be employed to estimate nonlinear functionals of a quantum state $\rho$ such as $\Tr(\rho^n)$ \cite{Ekert2002}, with a low-depth circuit.

Furthermore, methods for estimating gradient employed in variational quantum algorithms (VQA) also illustrate the correspondence between those two approaches for certain cases.
VQAs such as the VQE employ a parametrized quantum circuit $U(\bm{\theta})$ and classical optimizer, which minimizes a cost function $\mathcal{L}(\bm{\theta})$ by iteratively tuning the circuit parameter $\bm{\theta}$.
The cost is usually computed from expectation values of observables, therefore, the gradient of them can be a key ingredient for the optimization.
We can estimate the gradient in two ways.
Indirect and direct schemes have been proposed in Ref. \cite{Guerreschi2017} and Refs. \cite{Mitarai2018, Li2017b}, respectively.
The indirect method uses two different quantum circuit to estimate one element of the gradient.

These examples motivate us to further develop the methodology for replacing the indirect measurement with the direct measurement.
In this work, we describe the general protocol for such replacement.
The protocols for the Hadamard test involving a single controlled gate are given as Result 1 and 2.
Result 1 is a generalization of the method used in VQE, and Result 2 is a generalization of the destructive swap test which applies to general local unitary gates.
Finally, as Result 3, we describe a method to replace the Hadamard test involving multiple controlled gates.
It is a generalization of the method to estimate the gradient of observables with direct measurement, that is, we employ multiple quantum circuits to estimate the output of the Hadamard test.
The replacement can significantly reduce the depth of a quantum circuit and the accumulating effect of noise to the measured quantity.
Based on above results, we propose new methods to estimate higher order gradient including the metric tensor $g_{jk} = \frac{\partial \bra{\psi(\bm{\theta})}}{\partial \theta_j}\frac{\partial \ket{\psi(\bm{\theta})}}{\partial \theta_j}$ of variational quantum state $\ket{\psi(\bm{\theta})}$ and Hessian of an observable.
Specifically, the metric tensor is a key quantity in variational quantum simulations \cite{Li2016c,McArdle2018}.
Finally, we present a new protocol to measure multipoint correlator like out-of-time-order correlators (OTOC), which is an important quantity in a quantum many-body system as a possible measure of quantum chaos \cite{Rozenbaum2017, Swingle2017, Huang2017, Roberts2017}.

\noindent\textit{Hadamard test with one controlled gate -}
Let us first consider the case where the unitary $U$ in Fig. \ref{fig:Hadamard_test} is given by an exponential $U(\bm{\theta}) = e^{-i\theta G}$ of an Hermitian operator $G$ such that $G^2 = I$.
In this case, $\expect{Z}$ of the ancilla in the Hadamard test becomes $\braket{\psi|U|\psi} = \cos \frac{\theta}{2} - i \sin\frac{\theta}{2} \braket{\psi|G|\psi}$.
Hence the measurement of $\braket{\psi| G |\psi}$, which is just the expectation value of $G$, suffices to replace the Hadamard test.
We can evaluate it efficiently if $G$ can be decomposed as $G = \sum_{P\in \mathcal{P}} a_P P$, where $a_P \in \mathbb{R}$ and $\mathcal{P} = \{I, X, Y, Z\}^{\otimes n}$, with the polynomial number of terms with respect to the number of qubits.
More generally, the same applies if the gate $U$ itself can be decomposed into the sum of Pauli products consisting of the polynomial number of terms, which naturally include the previous case.
Therefore we have the following.

\textbf{Result 1}:
\textit{If the gate $U$ can be decomposed into the sum of Pauli products with the polynomial number of terms with respect to the number of qubits, the output of Fig.~\ref{fig:Hadamard_test}, $\braket{\psi|U|\psi}$, can be estimated with direct measurement by evaluating each Pauli terms.}

A prototypical example of the above result is the replacement of the phase estimation with direct measurements in the VQE.
The tradeoff of the protocol above is the time required to achieve the precision of $\epsilon$.
It scales as $O(1/\epsilon^2)$ in the direct approach and $O(1/\epsilon)$ in the indirect approach, i.e., the phase estimation.

Next, we describe another method for the case where the quantum gate $U$ is sufficiently local.
It is the generalization of the destructive swap test \cite{Garcia-Escartin2013}.
We say $U$ is $k$-local if $U$ can be decomposed into a tensor product of unitary matrices as $U=\bigotimes_q U_q$ and each $U_q$ acts on at most $k$-qubit.
With this definition, the result can be stated as follows.

\textbf{Result 2}:
\textit{Let $k$ be an integer such that $k=O(\poly(\log n))$, where $n$ is the number of qubits. For any $k$-local quantum gate $U$, it is possible to estimate $\bra{\psi}U\ket{\psi}$ up to the precision $\epsilon$ in time $O(k^2 2^k /\epsilon^2)$ without the use of the Hadamard test, with classical preprocessing of time $O(\poly(\log n))$.}

This result follows from the following reasons.
By the definition of $k$-local unitary matrix, $U$ can be decomposed into $U=\bigotimes_{q=1}^Q U_q.$
Let the number of qubits on which $U_q$ acts and eigenvalues of $U_q$ be $k_q$ and $\left\{\exp\left(i\phi_{q, m}\right)\right\}_{m=0}^{2^{k_q}-1}$, respectively, where $\phi_{q,m}\in [0,2\pi]$.
We denote the computational basis of each subsystem by $\ket{m_q}$ using integer $m_q = 0,\cdots,2^{k_q}-1$.
In this setting, $U_q$ is a $2^{k_q}\times 2^{k_q}$ matrix.
Classical computation can diagonalize each $U_q$  and obtain a unitary matrix $V_q$ such that $U_q = V_q^\dagger D_q V_q$, where $D_q = \sum_{m=0}^{2^{k_q}-1} e^{i\phi_{q, m}}\ket{m_q}\bra{m_q}$, in polynomial time to $n$ by the assumption $k=O(\poly(\log n))$.
Then,
\begin{align}
    &\bra{\psi}U\ket{\psi} \nonumber\\
    &=\bra{\psi}\left(\bigotimes_{q=1}^Q V_q^\dagger \sum_{m_q=0}^{2^{k_q}-1} \exp\left(i\phi_{q, m_q}\right)\ket{m_q}\bra{m_q}V_q\right)\ket{\psi} \\
    &=\sum_{m_1=0}^{2^{k_1}-1}\cdots \sum_{m_Q=0}^{2^{k_Q}-1} \left(\prod_{q=1}^Q\exp\left(i\phi_{q, m_q}\right)\right)\nonumber\\
    &\quad\left|\left(\bigotimes_{q=1}^Q\bra{m_q}\right)\left(\bigotimes_{q=1}^Q V_q\right)\ket{\psi}\right|^2.
\end{align}
Therefore, we can estimate $\bra{\psi}U\ket{\psi}$ by evaluating the probability of getting the result $\bigotimes_{q=1}^Q\ket{m_q}$ from the measurement of $\left(\bigotimes_{q=1}^Q V_q\right)\ket{\psi}$ in the computational basis.
More concretely, let the $j$-th measurement result be $m_q^{(j)}$ and the total number of measurement be $N$.
Then $\bra{\psi}U\ket{\psi}$ is estimated by $\bra{\psi}U\ket{\psi} \sim \frac{1}{N}\sum_{j=1}^N \prod_q \exp\left(i\phi_{q, m_q^{(j)}}\right)$.
The precision of the estimation $\epsilon = \left|\bra{\psi}U\ket{\psi} - \frac{1}{N}\sum_{j=1}^N \prod_q \exp\left(i\phi_{q, m_q^{(j)}}\right)\right|$ scales as $o(1/\sqrt{N})$ from the central limit theorem.
Since the number of gates to implement $V_q$ can scale as $O(k_q^2 2^{k_q} \poly(\log (k_q^2 2^{k_q})))$ in general \cite{Nielsen2010b}, the overall time for this protocol is $O(k^2 2^{k}\poly(\log (k^2 2^{k}))/\epsilon^2)$ with classical preprocessing for the diagonalization of each $V_q$ in time $O(\poly(\log n))$.



\begin{figure}
    \includegraphics[width=0.9\linewidth]{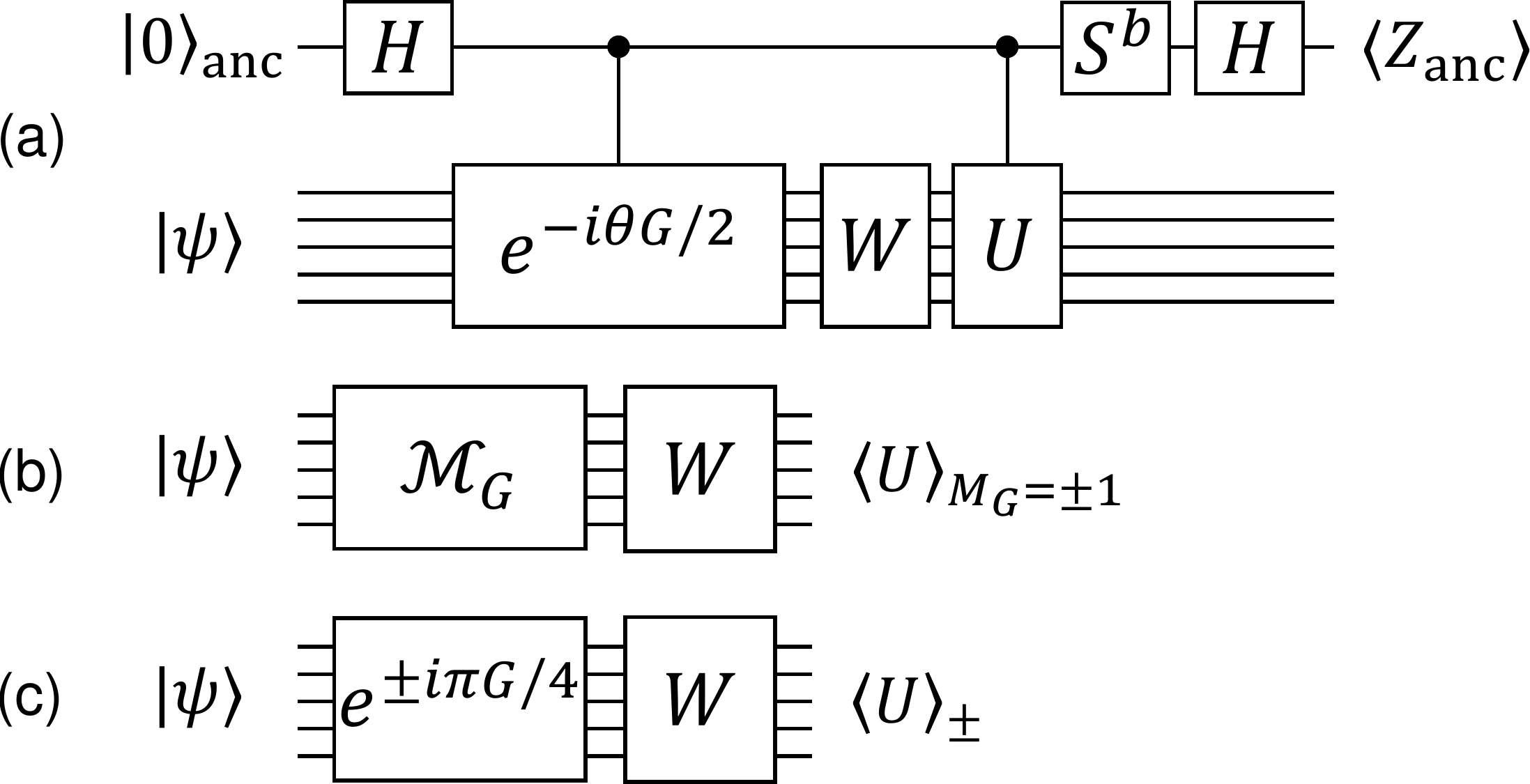}
    \caption{\label{fig:Hadamard_test_two_cgate_with_without_ancilla} (a) Hadamard test with two controlled gates. In the figure, $W$ is an arbitrary quantum gate. (b), (c) Quantum circuits to estimate the output of (a) with direct measurement. $\mathcal{M}_{G}$ is the projective measurement of $G$}
\end{figure}

\textit{Hadamard test with multiple controlled gates -}
Now we describe how to reduce an Hadamard test with multiple controlled gates to circuits without an ancilla qubit.
The protocol given below is for the Hadamard test with two controlled gates (Fig.~ \ref{fig:Hadamard_test_two_cgate_with_without_ancilla} (a)).
It is straightforward to generalize the method to the case of more than two controlled gates.
In the case of Fig.~\ref{fig:Hadamard_test_two_cgate_with_without_ancilla} (a), the measured quantity is $\braket{\psi| W^\dagger U W e^{-i\theta_1 G/2}|\psi}$.
If we assume $G^2=I$,
\begin{align}
    &\braket{\psi|W^\dagger U W e^{-i\theta_1 G/2}|\psi} \nonumber \\
    &= \cos\frac{\theta_1}{2} \braket{\psi|W^\dagger U W |\psi} 
    - i\sin\frac{\theta_1}{2} \braket{\psi|W^\dagger U W G|\psi}. 
\end{align}
The first term of the right hand side of the above formula is merely an expectation value of $U$ with respect to the state $W\ket{\psi}$, therefore, if $U$ satisfies one of the conditions mentioned in Results 1 and 2, we can evaluate it efficiently.
Even if it does not, the protocol which uses destructive swap test to measure $|\braket{\psi|U|\psi}|^2$ \cite{Subasi2018} can be utilized to estimate it using a quantum computer with $2n$ qubit.
For the second term, we present a method involving a projective measurement of $G
$, which we denote by $\mathcal{M}_G$.
When $G$ is a Pauli product, i.e. $G\in \{I,X,Y,Z\}^{\otimes n}$, $\mathcal{M}_G$ can be performed by first transferring $G$ to a single qubit $Z$ and then performing a measurement on the qubit nondestructively, for example with the dispersive readout of a superconducting qubit.
This protocol of $\mathcal{M}_G$ can, in principle, be generalized to $G$ such that the degeneracy of its eigenvalue $\pm 1$ is equivalent although such a circuit can be exponentially hard to construct.
On the other hand, for $G$ which does not satisfy the above condition, $\mathcal{M}_G$ requires an ancilla qubit due to the impossibility of transferring $G$ to a single qubit $Z$.
To estimate $\braket{\psi|W^\dagger U W G|\psi}$, we use the following four quantities for its estimation,
\begin{align}
    \expect{U}_{\pm} &= \braket{\psi|e^{\mp i\pi G/4}W^\dagger U W e^{\pm i\pi G/4}|\psi}, \\
    \expect{U}_{M_G = \pm 1} &= \frac{1}{4p(M_G=\pm 1)}\braket{\psi|(I\pm G)W^\dagger U W (I\pm G)|\psi}, 
\end{align}
where $p(M_G = \pm 1)$ is probability of getting the result $M_G = \pm 1$ by performing $\mathcal{M}_G$ on $\ket{\psi}$; $p(M_G = \pm 1) = \|\frac{1}{2}(I\pm G)\ket{\psi}\|^2$.
Figure~\ref{fig:Hadamard_test_two_cgate_with_without_ancilla} (b) and (c) show the quantum circuit to estimate them.
With these, $\braket{\psi|W^\dagger U W G|\psi}$ can be estimated by,
\begin{align}
    &\braket{\psi|W^\dagger U W G|\psi} \nonumber \\
    &= p(M_G=+ 1)\expect{U}_{M_G = + 1} - p(M_G=- 1)\expect{U}_{M_G = - 1}\nonumber \\
    &\quad + \frac{i}{2} (\expect{U}_- - \expect{U}_+). \label{eq:Hadamard_test_two_cgate}
\end{align}
Note that when $U$ is Hermitian, the first two terms corresponds to the real part, and the rest corresponds to the imaginary part of $\braket{\psi|W^\dagger U W G|\psi}$.
Therefore, we have the following.

\textbf{Result 3:}
\textit{Let $W$, $U$ and $G$ be unitary matrices. Suppose $U$ satisfies a condition of Results 1 and 2, and $G^2=I$. It is possible to estimate the output of the circuit in Fig. \ref{fig:Hadamard_test_two_cgate_with_without_ancilla} (a), $\braket{\psi|W^\dagger U W e^{-i\theta_1 G/2}|\psi}$, by using four quantum circuits in Fig. \ref{fig:Hadamard_test_two_cgate_with_without_ancilla} (b) and (c), and combining their output with Eq. (\ref{eq:Hadamard_test_two_cgate}). Especially, if eigenvalues $\pm 1$ of $G$ have equal degeneracy, the protocol can work without an ancilla qubit. Even if $U$ does not satisfy any condition of Results 1 and 2, }

We can extend the strategy of the methods described in the above sections to evaluate the output from the Hadamard test in the case where it has more than two controlled gates.
By the construction of this method, the number of terms that we need to measure grows exponentially to the number of controlled gates, while in practice a few controlled gates are enough as seen later.

\noindent\textit{Advantage -} Now let us discuss the possible merits of our protocols.
Result 1, 2 and 3 can provide a drastic reduction of depth of a quantum circuit.
The reduction is due to the fact that it is generally a hard task to make a controlled-$U$ gate for a given $U$.
For example, a Toffoli gate, which is a controlled-CNOT gate, requires at least six CNOT gates for its construction \cite{Shende2009}.
Along with the depth reduction, the disuse of an ancilla qubit is a nice advantage for qubit-limited, near-term quantum computers.
Result 3 has the same advantages as the above three, apart from the reduction of the effect of noise.
In the Hadamard tests of the type described in Fig.~\ref{fig:Hadamard_test_two_cgate_with_without_ancilla} (a), the ancilla qubit has to tolerate dephasing error during the gate $W$.
In contrast, the ancilla qubit used in the protocol of Result 3 does not have to endure the noise; it is measured projectively at $\mathcal{M}_G$.

\noindent\textit{Applications -} As the first application, we describe the direct measurement of derivatives of a parametrized quantum state.
In VQAs such as the VQE \cite{Peruzzo2013}, we employ a parametrized quantum circuit $U(\bm{\theta})$ and an input state $\ket{\psi_{\text{in}}}$ on an $n$-qubit quantum computer, and optimize the circuit parameter $\bm{\theta}$ with respect to an expectation value $\expect{A(\bm{\theta})} = \braket{\psi_{\text{in}}|U^\dagger(\bm{\theta})AU(\bm{\theta})|\psi_{\text{in}}}$ of an observable $A$.
Let us consider the case where the parametrized quantum circuit is constructed as $U(\bm{\theta}) = U_L(\theta_L)\cdots U_2(\theta_2)U_1(\theta_1)$ and each unitary $U_j(\theta_j)$ is generated by a Pauli product $P_j \in \{I,X,Y,Z\}^{\otimes n}$; $U_j(\theta_j) = \exp\left(-i\theta_j P_j/2\right)$.
We denote $U_k(\theta_k)\cdots U_j(\theta_1)$ by $U_{k:j}$.
In VQAs, we consider $A$ which can be decomposed into a sum of Pauli products, and therefore, without loss of generality, we assume $A \in \{I,X,Y,Z\}^{\otimes n}$.

\begin{figure}
    \includegraphics[width=0.9\linewidth]{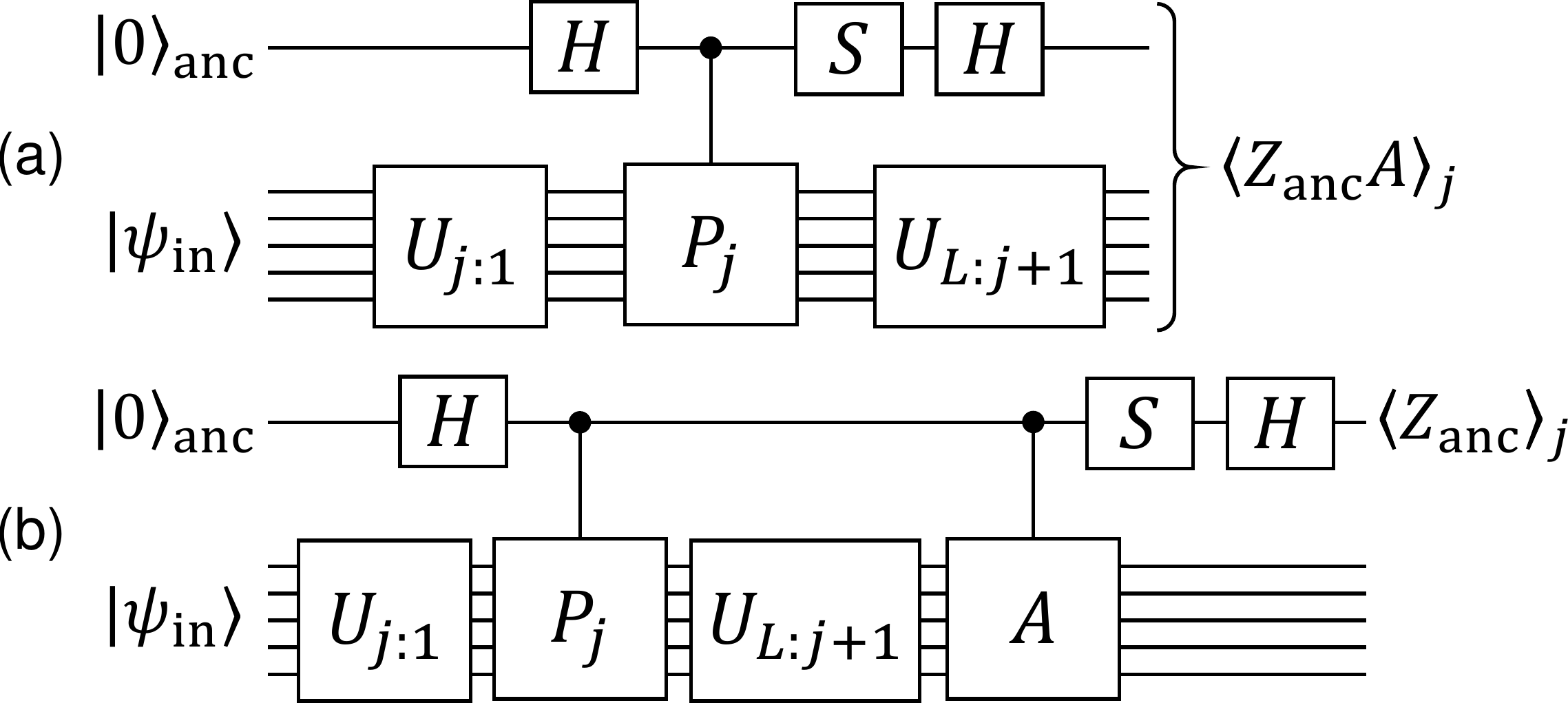}
    \caption{\label{fig:gradient_with_without_ancilla} Quantum circuit for the estimation of $\frac{\partial \expect{A(\bm{\theta})}}{\partial \theta_j}$. (a) Circuit with an ancillary qubit from \cite{Guerreschi2017}. $Z_{\mathrm{anc}}$ is the Pauli Z acting only on the ancillary qubit. The output of the circuit, $\expect{Z_{\text{anc}}A}_j$, corresponds to the gradient as $\expect{Z_{\mathrm{anc}}A} = -\frac{\partial \expect{A(\bm{\theta})}}{\partial \theta_j}$. (b) Quantum circuit which has equivalent output as (a); $\expect{Z_{\text{anc}}A}_j = \expect{Z_{\text{anc}}}_j$.}
\end{figure}

In Fig.~\ref{fig:gradient_with_without_ancilla} (a), we show a quantum circuit to evaluate the analytic gradient of an expectation value of an observable $A$ as presented in Ref.~\cite{Guerreschi2017}.
The circuit of Fig.~\ref{fig:gradient_with_without_ancilla} (b) measures an equivalent quantity, and this circuit is equivalent to that one in Fig.~\ref{fig:Hadamard_test_two_cgate_with_without_ancilla} (a) when we replace $e^{-i\theta G/2}$ with $P_j$, $U$ with $A$.
Since we assumed $A$ is a Pauli product, it satisfies the condition in Result 1.
Therefore, Result 3 can be employed to measure the gradient.
Note that the gradient of $A$ is
\begin{equation}
\frac{\partial A}{\partial \theta_j} = \text{Im}\left(\braket{\psi_{\text{in}}|U^\dagger_{j:1} U_{L:j+1}^\dagger A U_{L:j+1} P_j U_{j:1}|\psi_{\text{in}}}\right),
\end{equation}
which only involves the last two terms of Eq.~(\ref{eq:Hadamard_test_two_cgate}), and hence we do not need to evaluate the circuit in Fig.~\ref{fig:gradient_with_without_ancilla} (b).
This method can easily be extended to evaluate the Hessian.

\begin{figure}
    \includegraphics[width=\linewidth]{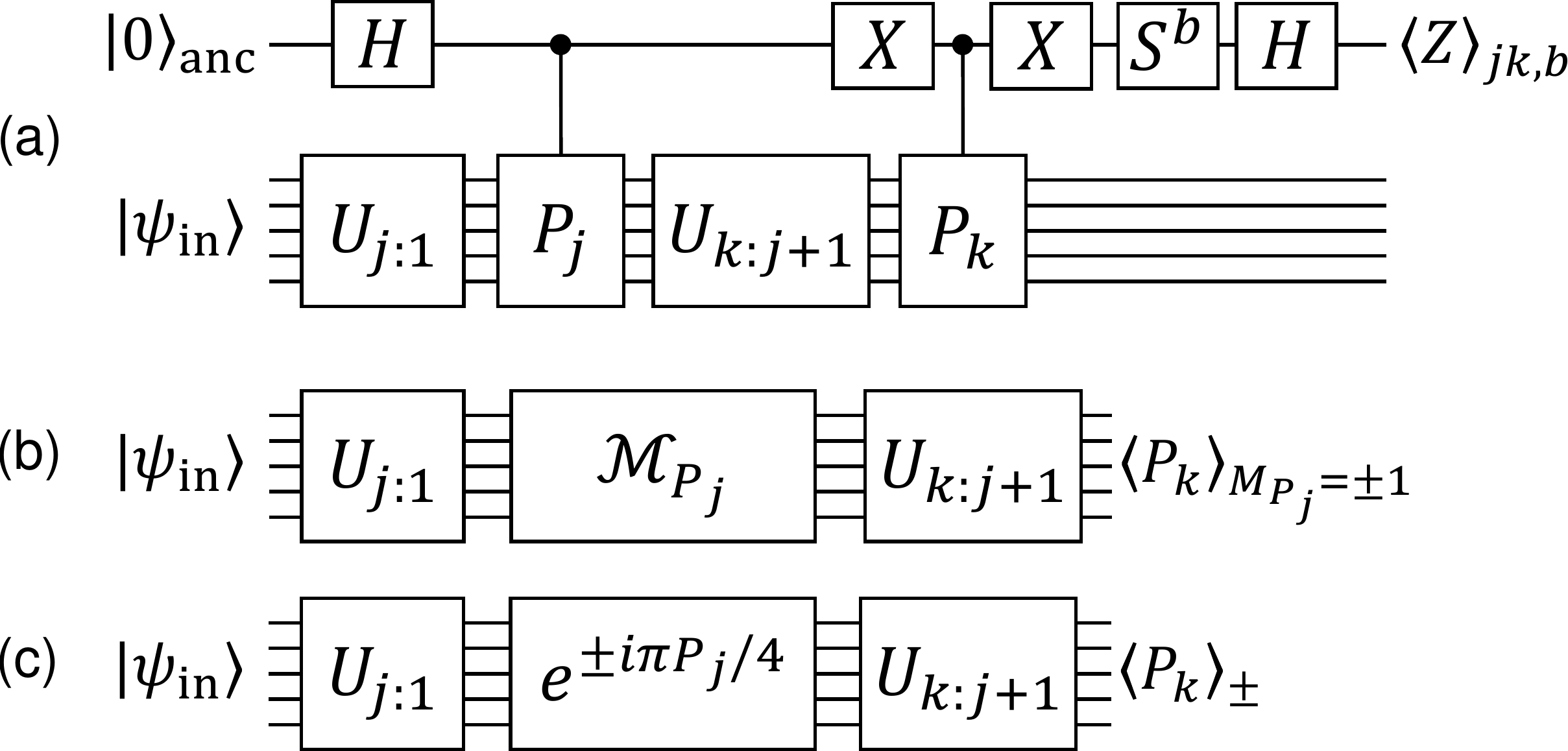}
    \caption{\label{fig:metric_tensor} Quantum circuit for the estimation of the real and imaginary part of the metric tensor $g_{jk}$. (a) Indirect method from Refs. \cite{Li2016c, McArdle2018}. $b\in \{0,1\}$. When $b=0$, $\expect{Z_{\mathrm{anc}}}_{jk,0}=4\mathrm{Re}(g_{jk})$ and when $b=1$, $\expect{Z_{\mathrm{anc}}}_{jk,1} = 4\mathrm{Im}(g_{jk})$. (b) Direct method to estimate the real part of $g_{jk}$. (see Eq.~(\ref{eq:real_metric_without_ancilla_explicit}).) (c)Direct method to estimate the imaginary part of $g_{jk}$. (see Eq.~(\ref{eq:imaginary_metric_without_ancilla_explicit}).)}
\end{figure}

The metric tensor  $g_{jk} = \frac{\partial \bra{\psi(\bm{\theta})}}{\partial \theta_j}\frac{\partial \ket{\psi(\bm{\theta})}}{\partial \theta_j}$ of a variational quantum state $\ket{\psi(\bm{\theta})} = U(\bm{\theta})\ket{\psi_{\text{in}}}$ can be measured in the same manner.
This quantity is the key for executing variational quantum simulations.
Specifically, the imaginary and the real part of $g_{jk}$ are employed for the simulation of real  \cite{Li2016c} and imaginary time \cite{McArdle2018} evolutions, respectively.
A quantum circuit for its evaluation from Refs. \cite{Li2016c, McArdle2018} is shown as Fig.~\ref{fig:metric_tensor}~(a).
The explicit expression for $g_{jk}$, when $k>j$, can be written as:
\begin{align}
    g_{jk}
    &= \frac{1}{4}\braket{\psi_{\mathrm{in}}|U_{j:1}^\dagger P_j U_{k:j+1}^\dagger P_k U_{k:1}|\psi_{\mathrm{in}}}. \label{eq:metric_tensor} 
\end{align}
Figure~\ref{fig:metric_tensor} (a) shows the quantum circuit for the indirect measurement of $g_{jk}$.
Again, by Result 3, This circuit can be replaced with the ones in Fig.~\ref{fig:metric_tensor} (b) and (c).
An explicit expression is:
\begin{align}
    \mathrm{Re}(g_{jk})
    = \frac{1}{4}&\left(p(M_{P_j}=+1)\expect{P_k}_{M_{P_j} = +1}\right. \nonumber
    \\ &\quad \left. - p(M_{P_j}=- 1)\expect{P_k}_{M_{P_j} = -1}\right), \label{eq:real_metric_without_ancilla_explicit}\\
    \mathrm{Im}(g_{jk}) = 
    -&\frac{\expect{P_k}_+ -\expect{P_k}_-}{8}. \label{eq:imaginary_metric_without_ancilla_explicit}
\end{align}
Figure~\ref{fig:metric_tensor} (a) differs from Fig.~\ref{fig:Hadamard_test_two_cgate_with_without_ancilla} (a) with two additional $X$ gates on the ancilla.
The consequence of this is the sign flip in the imaginary part (compare Eq.~(\ref{eq:imaginary_metric_without_ancilla_explicit}) and (\ref{eq:Hadamard_test_two_cgate})).

\begin{figure}
    \includegraphics[width=\linewidth]{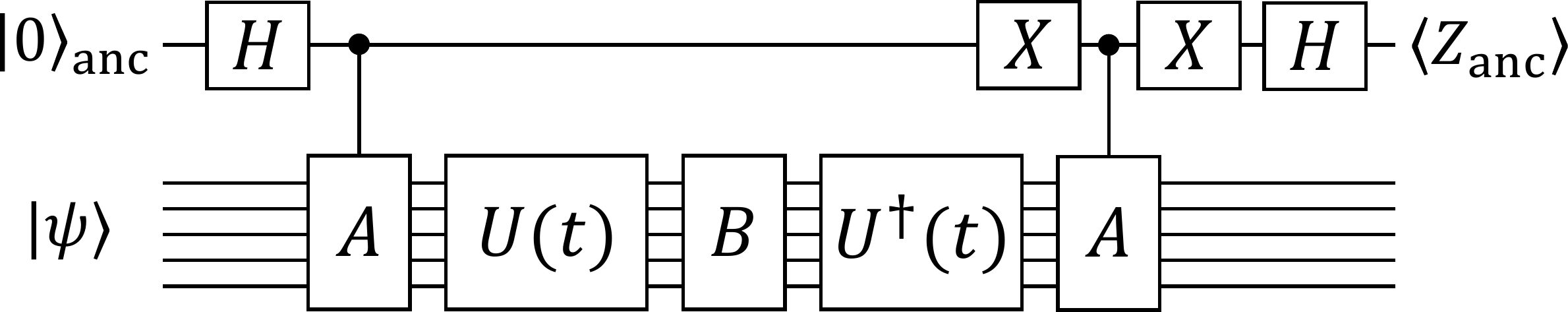}
    \caption{\label{fig:OTOC} Indirect approach to measure OTOC of oprators $A$ and $B$ from Ref.~\cite{Swingle2016}. In the figure, $U(t)=e^{-iHt}$.}
\end{figure}

Next, we propose a new method to estimate the OTOC on quantum computers.
The OTOC $F(t)$ at time $t$ is defined with two non-commuting operator $A$ and $B$ and a system Hamiltonian $H$ as $F(t) = \expect{B^\dagger(t) A^\dagger B(t) A}$, where $B(t) = e^{iHt}Be^{-iHt}$.
It is an important quantity in quantum many-body physics measuring how chaotic a given quantum system is \cite{Rozenbaum2017, Swingle2017, Huang2017,Roberts2017}.
Ref.~\cite{Swingle2016} has proposed the circuit  to evaluate $F(t)$ as shown in Fig. \ref{fig:OTOC}.
If we assume $A^2 = I$, the circuits in Fig.~\ref{fig:Hadamard_test_two_cgate_with_without_ancilla} (b) and (c) and Eq.~(\ref{eq:Hadamard_test_two_cgate}) can be applied directly, with the sign flip of the imaginary part as the consequence of the $X$ gates performed on the ancilla qubit.
More concretely, to evaluate $F(t)$, we replace $W$ in Eq.~(\ref{eq:Hadamard_test_two_cgate}) with $U^\dagger(t)B U(t)$, $U$ and $G$ with $A$.
This method can readily be extended to the measurement of higher order correlators.

\noindent\textit{Conclusion -}
We provided general protocols to replace the indirect measurement, especially, the Hadamard test, with the direct measurement.
The proposed methods to replace the Hadamard test provides means to evaluate analytical gradient, metric tensor, Hessian, and even higher order derivatives with direct measurement for parameter tuning in variational quantum algorithms.
It can also be applied for the estimation of OTOC.
The presented protocols can significantly reduce the depth of a quantum circuit, and consequently, are important subroutines for quantum algorithms, especially for those of NISQ devices.

\begin{acknowledgements}
    KM thanks METI and IPA for its support through MITOU Target program.
    KF is supported by KAKENHI No.16H02211, JST PRESTO JPMJPR1668, JST ERATO JPMJER1601, and JST CREST JPMJCR1673.
    This work is supported by MEXT, Q-LEAP.
\end{acknowledgements}

\bibliographystyle{apsrev4-1}

\end{document}